# Deepfake News Detection: A Multimodal Framework Integrating LipNet, DeepSpeech and ResNET for Enhanced Audio-Visual Analysis

Ameena Khan[1,3], Muhammad Ahsan Aziz [2], Muhammad Junaid Asif [3], Naeem Akhter [1] and Rana Fayyaz Ahmad [3]
[1] Department of Computer and Information Sciences, Pakistan Institute of Engineering and Applied Sciences (PIEAS) Islamabad, Pakistan
[2] Department of Computer Science, Quaid-i-Azam University (QAU), Islamabad, Pakistan
[3] Artificial Intelligence Technology Centre (AITeC), National Centre for Physics (NCP), Islamabad 44000, Pakistan

***Abstract*—Deepfake news refers to AI-generated (or AI manipulated) multimedia content intentionally generated to deceive audiences by manipulating the facial expressions, or speech while maintaining the realistic appearance. The rapid progress of generative AI has made the synthesis of highly realistic fake videos and cloned voices widely accessible, posing a serious threat to the authenticity of digital news media. This paper presents a multi-modal framework that discerns the authenticity of video content by jointly exploiting audio and visual cues, thereby addressing the challenge of detecting the deepfake videos. We proposed a framework that involves features extraction from lip movements, audio content and video frames. Lip movements and speech content are encoded using the LipNet and DeepSpeech2 models, while facial features are extracted by leveraging the use of BlazeFace and represented with ResNet18. The extracted feature vectors are concatenated into a holistic video representation and classified with an ensemble of machine learning and deep learning models, including Random Forest (RF), Multi-layer Perceptron (MLP) and Long Short-Term Memory (LSTM) networks. Extensive experiments performed on the FakeAVCeleb dataset shows that the proposed approach attains an accuracy of 94% using augmented audio features, outperforming a state-of-the-art multimodal ensemble baseline. The results confirm the robustness and practical potential of the proposed framework for deepfake news detection.**

## I. Introduction

In the realm of AI, synthetic media refers to manipulated or artificially generated audios, images and videos that convincingly imitate real content. Deepfake technology superimposes the likeness of one individual onto another and can create convincing videos depicting a target person engaged in actions or speech they never performed. Deepfake news consists of these manipulated videos to deceive the audience and is widely spread across social media and news platforms. The mechanism behind deepfake creation relies on sophisticated deep-learning frameworks, notably autoencoders and Generative Adversarial Networks (GANs), which are widely used in computer vision, machine learning and deep learning to analyze and synthesize facial expressions and gestures [1]. With recent advances in artificial intelligence, manipulated videos in diverse social and political contexts present a significant threat to society and can be exploited for malicious purposes. **Fig. 1** depicts the process of deep-fake generation base on encoder-decoder architecture, where facial images are encoded into latent embeddings and decoded to generate realistic manipulated faces.

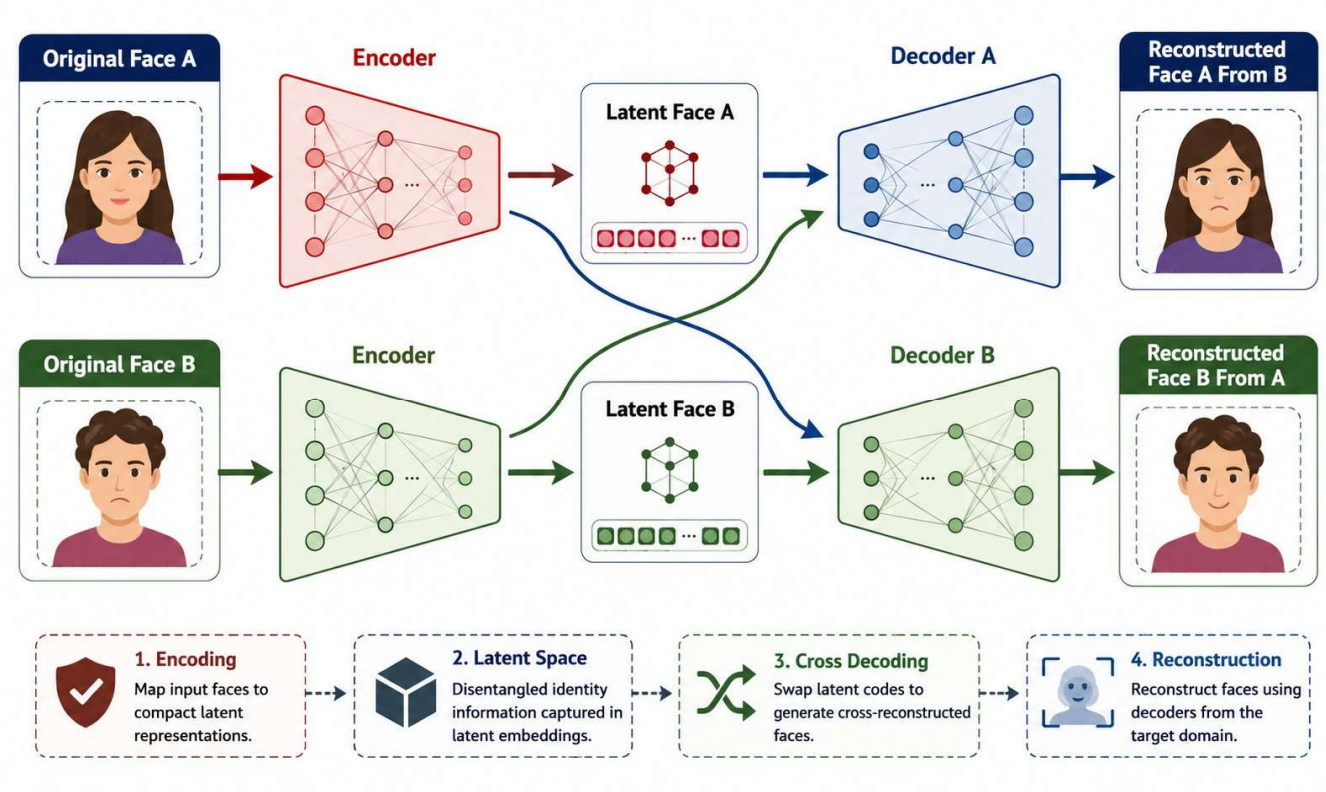

Fig. 1. Overview of the deepfake generation process based on an autoencoder architecture. Original facial images from two identities are encoded into latent representations using a shared encoder and subsequently reconstructed through identity-specific decoders, enabling cross-face synthesis for deepfake generation.

The growth of synthetic multimedia content, encompassing both audio and video, is simultaneously transforming several industries such as e-commerce, healthcare and historical documentation. The dual nature of this media is, however, a matter of concern: while it can stimulate innovation and improve user experience, it also carries the risk of abuse and exploitation. Although significant strides have been made in enhancing detection accuracy, the continuous evolution of deepfake generation techniques underscores the persistent demand for a robust deepfake news detection system with enhanced performance metrics. The complexity of the problem stems from the immense volume of content circulating online, comprising both genuine information and disinformation, and from the critical task of ascertaining the veracity of the news and the identity of the newscaster, whether in real-time broadcasts or pre-recorded segments.

In contrast to the predominant body of work that addresses deep-fake detection from a single modality—either video or audio—this research proposes a comprehensive multimodal news detection model that elevates the efficacy of detection through refined representation learning. The distinctiveness of the approach lies in coupling semantic audio features extracted from lip movements ***(LipNet)*** and speech-to-text content (DeepSpeech2) with visual neural features, and in evaluating the covariance between these semantic features so that only the most informative sub-networks are retained. The framework is organized into discrete sub-networks, each autonomously contributing to feature extraction, whose outputs are concatenated before classification; this design balances accuracy with computational efficiency, an aspect that is essential for real-world deployment but is frequently overlooked in prior multimodal pipelines.

The principal objective of this study is to propose an improved method of differentiating fake content from real content. Different facial and acoustic features are detected using machine-learning and deep-learning models, and the retrieved features are trained on large datasets so that the entire model is learned end-to-end. Because high-grade deep-fake videos and images defeat many previously suggested algorithms, the goal is to propose an advanced and optimized system that can correctly categorize manipulated videos and audio. To this end, the model processes both visual neural features and audio spectral features, harnessing transfer learning from cutting-edge pre-trained models, and introduces a semantic-based feature analysis technique that augments detection capability while exhibiting robust generalization across diverse datasets. In particular, this work investigates how audio and visual modalities can be jointly represented to reliably distinguish authentic news videos from manipulated ones, which semantic features contribute most to detection performance, how the choice of classifier affects accuracy on a multimodal feature vector, and what impact data augmentation has on detection performance for an imbalanced deepfake dataset.

The main contributions of this paper can be summarized as follows. First, a multimodal and ensemble framework is proposed that combines recent and promising deepfake detection approaches across audio and visual modalities for robust detection generalization. Second, transfer learning is exploited from four cutting-edge pre-trained models—***LipNet*** and ***DeepSpeech2*** for audio, ***BlazeFace*** and ***ResNet18*** for faces—whose semantic features are concatenated into a single holistic representation. Third, a semantic-based feature analysis technique is introduced in which each sub-network is independently assessed and only the most effective outputs are retained, jointly optimizing accuracy and computational efficiency. Finally, a comprehensive comparison of machine-learning and deep-learning classifiers is conducted on the ***FakeAVCeleb*** dataset, achieving 94 percent accuracy and surpassing a state-of-the-art ensemble multimodal baseline.

A high-level overview of the proposed work is presented in **Fig. 2**. This research analyses several key attributes of deepfake multimedia content, spanning audio waveforms, two-dimensional facial landmarks and mouth regions. These complementary cues form the basis of the multimodal representation: the speech signal is processed by DeepSpeech2, the cropped mouth regions by LipNet and the localized faces by ResNet18, after which the resulting feature vectors are concatenated and classified by an ensemble of learners to produce the final real/fake decision.

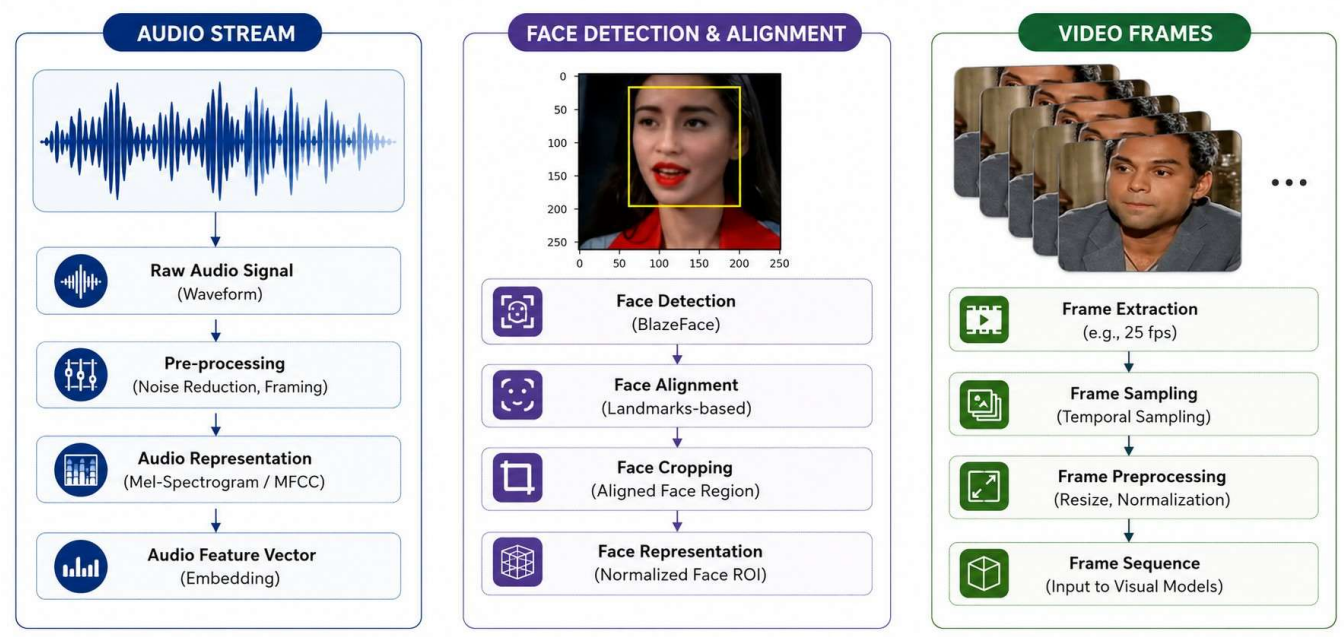

Fig. 2. Overview of the proposed multimodal feature extraction framework, illustrating the three complementary modalities analyzed for deepfake news detection: speech signals (audio waveforms), facial appearance (two-dimensional facial regions), and lip-motion information (mouth regions).

The remainder of this paper is organized as follows. Section II reviews existing video, audio and multimodal deepfake detection approaches and identifies the research gap that motivates this work. Section III details the proposed multi-modal methodology, including the audio and visual feature extractors and the classification models. Section IV describes the datasets, preprocessing and implementation details together with the evaluation metrics. Section V presents and discusses the experimental results, and Section VI concludes the paper and outlines directions for future work.

## II. Literature Review

The rapid advancement of deep learning and deepfake technology has ushered in a new era of both promise and peril. A poignant illustration is a lip-sync deepfake of former U.S. President Barack Obama produced by Jordan Peele, in which Obama appeared to make disparaging remarks he never uttered [2]. During the 2017 Indian election campaign a deepfake video propagated across messaging platforms, reaching an estimated fifteen million individuals and potentially swaying their voting decisions [3]. Beyond political manipulation, deepfake audio has been used to defraud the CEO of a UK-based energy company, causing a substantial financial loss [4].

### *A. Detection of Video Deepfakes*

Convolutional Neural Networks (CNNs) are ubiquitous in this domain owing to their feature-extraction ability [5], and transfer learning with architectures such as Xception-Net has been employed to identify manipulated images [6]. Afchar et al. proposed MesoNet, which scrutinises mesoscopic facial attributes using a shallow network [7]. To capture frame-level temporal inconsistencies, Guera and Delp combined CNN feature extraction with Long Short-Term Memory (LSTM) networks for sequence prediction [8]. Ciftci et al. introduced ***FakeCatcher,*** which exploits biological signals concealed in video content [9], while Li et al. detected eye-blinking irregularities and warping artifacts [10], [11], and later examined blending boundaries with the Face X-ray technique [12]. Many of these approaches rely on pixel- or image-level ground truth, which limits their applicability to complex real-world forgeries.

### *B. Detection of Audio Deepfakes*

Several approaches counter audio spoofing in automated speaker verification by exploring acoustic aspects [13], [14]. Multimodal techniques capitalize on Mel-Frequency Cepstral Coefficients (MFCCs) as acoustic features together with facial visual cues [15]–[17]. Traditionally, handcrafted Linear Frequency Cepstral Coefficients (LFCC) have been integrated with the Gaussian Mixture Model (GMM) for robust spoofing detection [18], whereas more recent research embraces deep learning, harnessing CNNs and Recurrent Neural Networks (RNNs) to scrutinize synthetic audio [19].

### *C. Multimodal Deepfake Detection*

Joint audio-visual representation learning remains comparatively under-explored. The first multimodal network for audio-visual deepfake detection faces several limitations that contribute to its lower accuracy, since the selection and integration of multimodal features may not effectively capture the complex spatial, spectral and temporal inconsistencies in deep-fake videos [20]. An emotion-based audio-visual method was proposed but evaluated mainly on publicly accessible datasets, omitting the jointly manipulated content of FakeAVCeleb [21]. A dissonance-based detector scrutinized the discordance between audio and visual constituents but was not assessed on jointly manipulated content [22]. Comprehensive evaluations of unimodal and multimodal detectors revealed challenges in robust generalization on test data [23]. Ensemble transformers integrating audio and visual modalities tend to overlook vital audio-visual cues [24], and ensembling audio, video and audio-visual models improves performance at the cost of redundancy across the three modalities [25].

In summary, although deepfake detection has progressed rapidly, the reviewed literature exposes three persistent gaps. First, most detectors remain unimodal, concentrating on either video or audio, and therefore miss the cross-modal inconsistencies that characterize modern deepfake news. Second, the multimodal methods that do combine audio and video frequently rely on handcrafted acoustic features or heavy ensembles that overlook fine-grained lip-sync cues and generalize poorly to unseen manipulation techniques and to jointly manipulated benchmarks such as ***FakeAVCeleb***. Third, very few approaches jointly optimize detection accuracy and computational efficiency, which is critical for screening the large volume of news content circulating in real time. These gaps collectively establish the need for a detector that fuses transfer-learned audio and visual semantic features in a selective, efficiency-aware manner. The methodology presented next addresses this need by coupling LipNet and DeepSpeech2 audio features with BlazeFace and ResNet18 visual features and retaining only the most informative sub-networks before ensemble classification.

## III. Proposed Methodology

The proposed model is built to succeed in detection generalization: it effectively distinguishes real content from altered content using a flexible multimodal and ensemble network that combines recent and promising deepfake detection approaches. Both audio and visual features are used, harnessing transfer learning from the cutting-edge audio models LipNet [26] and DeepSpeech2 [27] and the video models BlazeFace [28] and ResNet18 [29]. The topology comprises several sub-networks, each separately assessed for its effect on overall detection performance; using the semantic features extracted from these models, the covariance between them is examined. Both accuracy and computational efficiency are considered, and only the sub-networks that show the greatest effectiveness are kept. The feature vectors derived from the chosen sub-networks are then concatenated and fed into a final classifier. The overall flow of the proposed technique is shown in **Fig. 3**.

### ***A. Audio Feature Extraction***

***1) LipNet Model:*** LipNet [26] is an end-to-end lipreading architecture that transforms variable-length sequences of video frames into text sequences. Its basic principle is that spoken language can be understood from visual cues alone, notably from the precise movements of a speaker's lips. The mouth region is located and cropped in each frame as the area of interest, and these mouth frames form the main input. LipNet processes the frames in order, extracting both spatial and temporal characteristics of the lip movements. In this work only the features from the extracted mouth regions are used. The architecture is shown in **Fig. 4**.

***2) DeepSpeech2 Model:*** DeepSpeech2 [27] is a state-of-the-art Automatic Speech Recognition (ASR) model that converts spoken language into text. The raw audio is first converted into Mel spectrograms, a compact time–frequency representation of the signal. Convolutional and recurrent neural networks then process the spectrograms: CNNs capture spatial patterns while RNNs model the sequential nature of speech. A Connectionist Temporal Classification (CTC) objective aligns the acoustic features with the textual representations and copes with varying speaking rates and word lengths. At inference time the input audio is segmented into one-second spectrograms from which informative features are extracted, as depicted in **Fig. 5.**

### ***B. Face Feature Extraction***

***1) BlazeFace Model:*** BlazeFace [28] provides strong facial detection and landmark localization and is a lightweight,

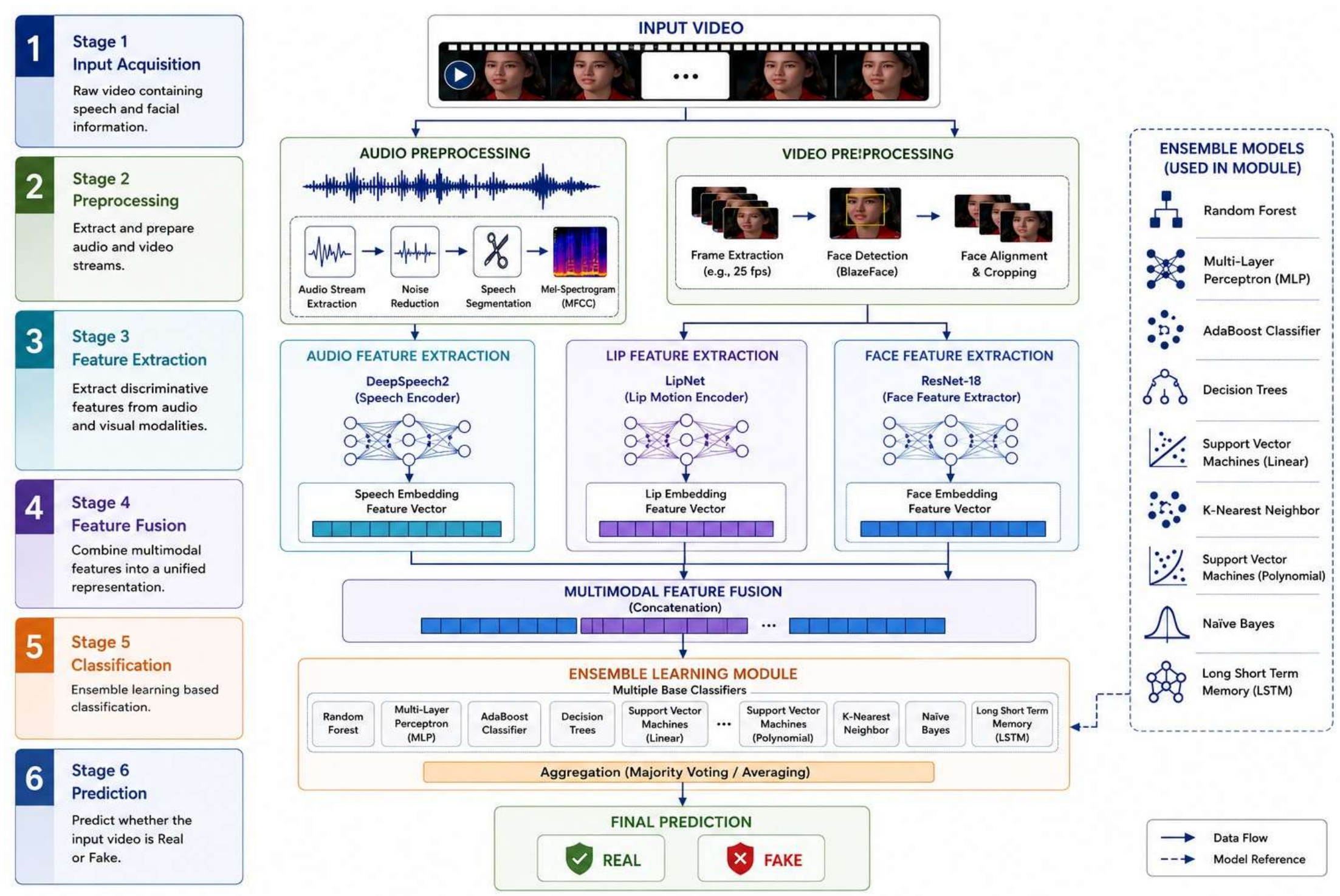

Fig. 3. Classification stage of the proposed framework. The concatenated feature vector is fed to an ensemble of classifiers including Random Forest, MLP, AdaBoost, Decision Trees, SVM, K-Nearest Neighbor, Na¨ıve Bayes and LSTM.

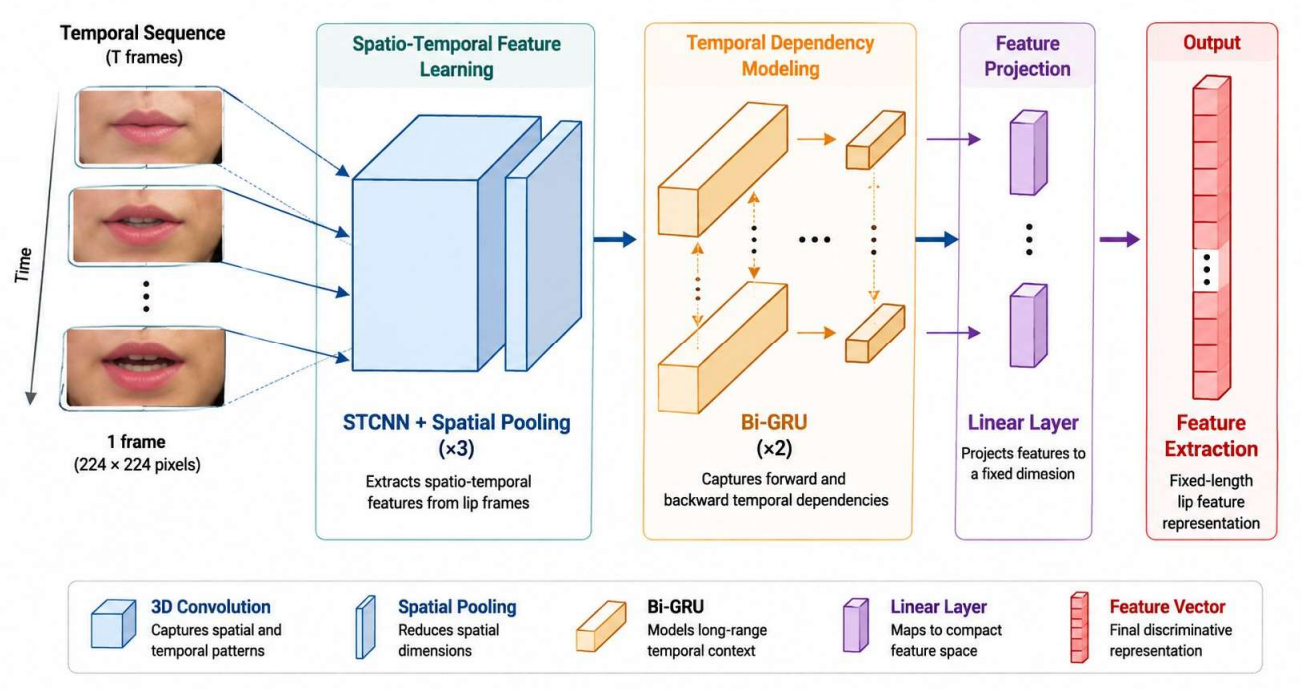

Fig. 4. Architecture of the LipNet feature extraction module. Sequential mouth-region frames are processed through spatiotemporal convolutional layers with spatial pooling, followed by bidirectional GRU layers and a linear projection layer to learn discriminative lip-motion feature representations.

computationally efficient model designed for mobile GPUs. It detects faces within individual frames even when faces vary in size, position or illumination, and pinpoints six essential facial landmarks—left eye, right eye, mouth, nose, left ear and right ear. The detected face is cropped from the surrounding frame and resized to a standard size of 299 × 299 pixels to maintain consistency for subsequent stages. Accurate landmark localization is crucial because deep-fake manipulations frequently entail small adjustments to these facial structures. The output is illustrated in Fig. 6.

***ResNet18 Model:*** ResNet18 [29], a member of the Residual Network family developed to address vanishing gra-dients in very deep networks, is given the detected faces. Its main technological advancement is the residual block, whose shortcut (skip) connections allow gradients to flow more naturally during training. ResNet18 was originally created for object detection and image classification, but owing to its exceptional feature-extraction capability it is used here to extract distinguishing features from the detected faces, capturing the basic traits and patterns specific to each face.

### C. Classification Models

The combined features of each extractor are concatenated and passed to a suite of machine-learning and deep-learning classifiers for comparison, as shown in **Fig. 3.**

*1) Random Forest:* Random Forest is a powerful ensemble learning method that builds several decision trees during training and combines their results. Each tree is built from a subset of the training data and a randomly chosen subset of features, and the resulting diversity makes the model robust and able to generalize well. Its ability to handle high-dimensional feature spaces makes it well suited to the concatenated feature vector produced by the DeepSpeech2, LipNet and ResNet18 extractors, from which it predicts whether a video is real or fraudulent.

***2) Multi-layer Perceptron:*** A Multi-layer Perceptron (MLP), a feedforward artificial neural network, is included for its ability to identify complex patterns. The model used here has an input layer, two hidden layers and an output

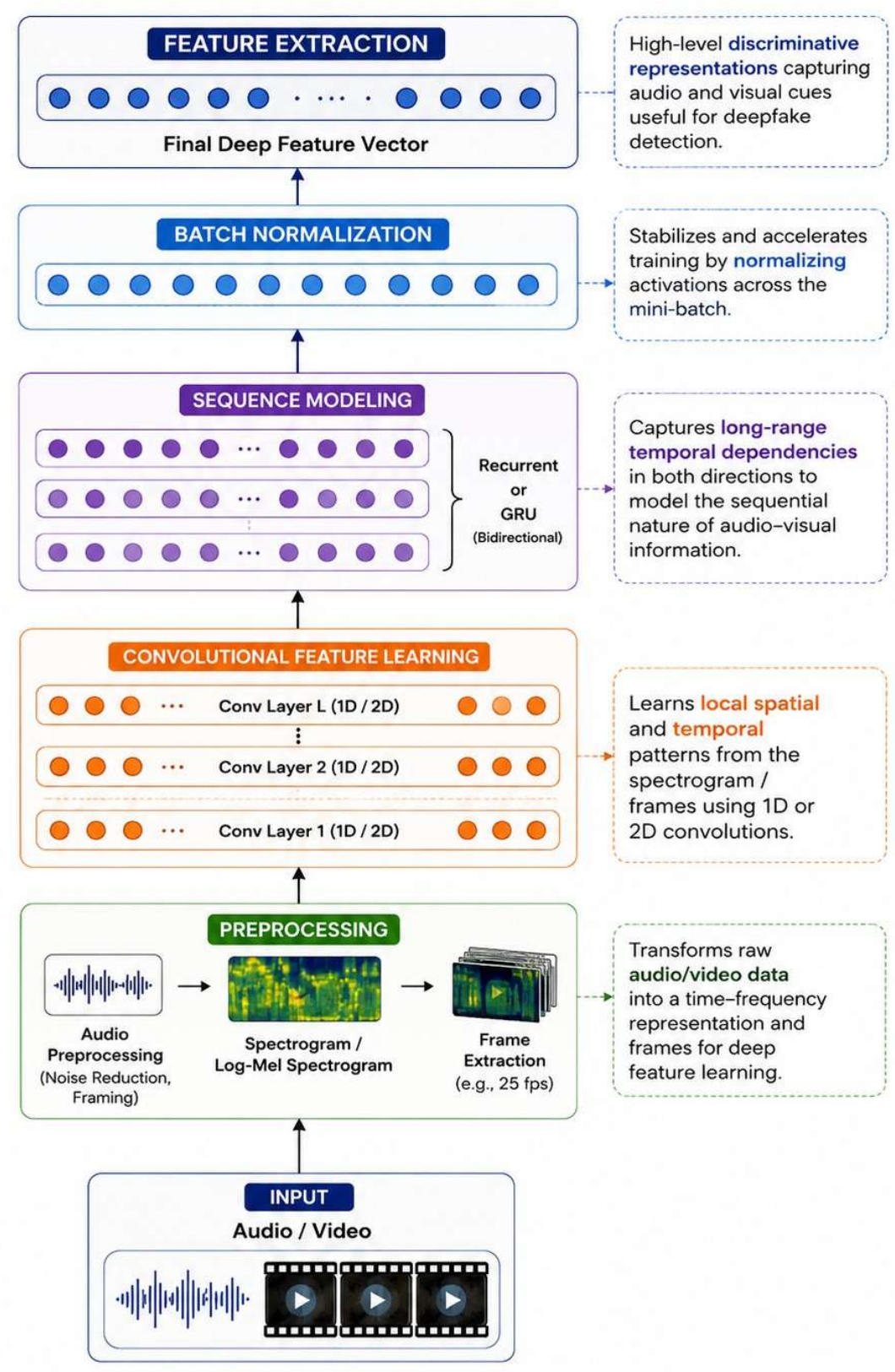

Fig. 5. DeepSpeech2 model: invariant convolution layers, batch normalisation and bidirectional recurrent/GRU layers used for audio feature extraction.

```
blaze detection: 0.009867429733276367
time: 0.03527355194091797
```

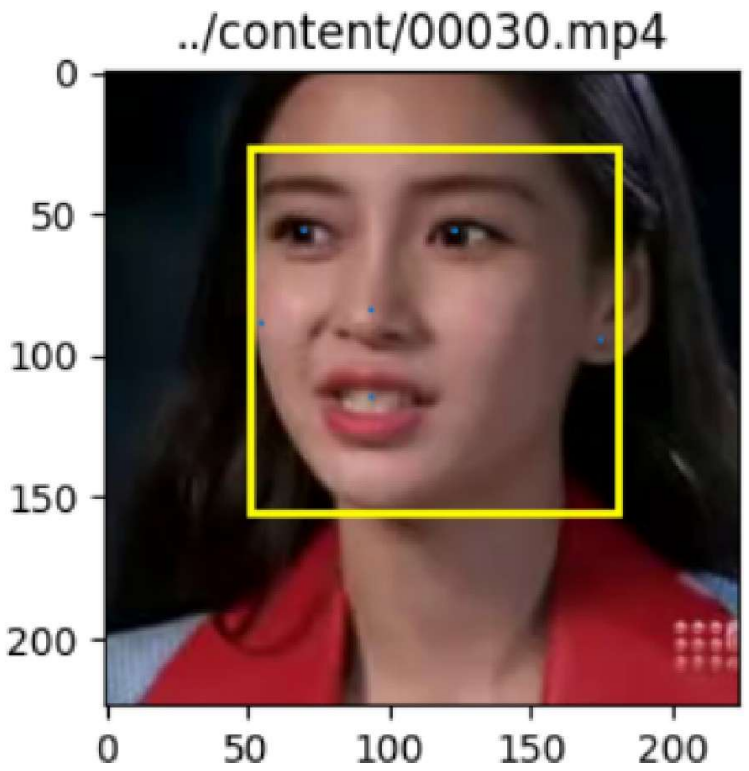

Fig. 6. BlazeFace model output: a bounding box and six facial landmarks localized on the detected face.

layer. During training, backpropagation adjusts the weights and biases to map the concatenated feature vector to the correct class label, minimizing a cross-entropy loss with gradient-descent optimization.

*3) Long Short-Term Memory:* An LSTM, a type of recurrent neural network specialized for sequence data, is adapted to the structure of the input. Because the features of a video do not exhibit temporal dependencies between frames, each video is treated as a separate sequence with a single time step; the sequence length is therefore set to one, effectively transforming the LSTM into a feedforward network that still captures complex relationships within the feature space.

### D. *Fine-Tuning*

Fine-tuning programs a deep-learning procedure with weights from previously trained transfer-learning models. Because the approach reuses valuable information from existing detection models, it drastically reduces the time needed to develop and execute a new detector. The generic convolutional features remain useful for categorizing visuals, while the final classification layer is specialized to the real/fake task of this work.

## IV. Dataset and Implementation

### A. *Datasets*

Two diverse and well-known datasets that contain both high-quality audio and video information are used to ensure that the model is adaptive: the **FakeAVCeleb** dataset [30] and the Deepfake Detection Challenge (DFDC) dataset [31].

**FakeAVCeleb** [30] is a multimodal deepfake dataset that includes synthesized lip-synced fake audio in addition to deepfake videos. Its 500 real videos are sampled from the VoxCeleb2 corpus of celebrity interviews with an average clip duration of about 7.8 seconds, and roughly 19,500 fake videos are generated from them, yielding a highly imbalanced set of some 20,000 clips. To address racial bias, the real videos are drawn equally from four ethnic groups (Caucasian, Black, South Asian and East Asian) with a balanced gender ratio. The dataset is organized into four audio-visual categories—real-audio/real-video, fake-audio/real-video, real-audio/fake-video and fake-audio/fake-video—so that both unimodal and jointly manipulated forgeries are represented, which makes photorealistic lip-synced fakes its principal challenge.

The DFDC dataset [31] is the largest publicly accessible collection of face-swap videos, comprising 128,154 ten-second clips recorded from 3,426 paid, consenting actors under varied lighting, pose and background conditions. The forgeries are produced with eight Deepfake, GAN-based and non-learned methods, and the set is imbalanced towards fake content; its synthesized audio is not synchronized with the visuals, and the manipulation labels are intentionally undisclosed, making generalization to unseen techniques its defining challenge.

### B. *Preprocessing and Augmentation*

A custom dataset is created by preprocessing each video into five or more real instances, which enables training and evaluation of the neural networks. Data augmentation is then applied to generate diverse renditions of the original data and to mitigate the class imbalance by generating additional minority-class instances. As shown in Fig. 7, augmentation is performed by sliding a window over one-second segments of the video. The resulting preprocessed FakeAVCeleb set contains 50,000 real and 100,000 fake videos for a total of 150,000 videos.

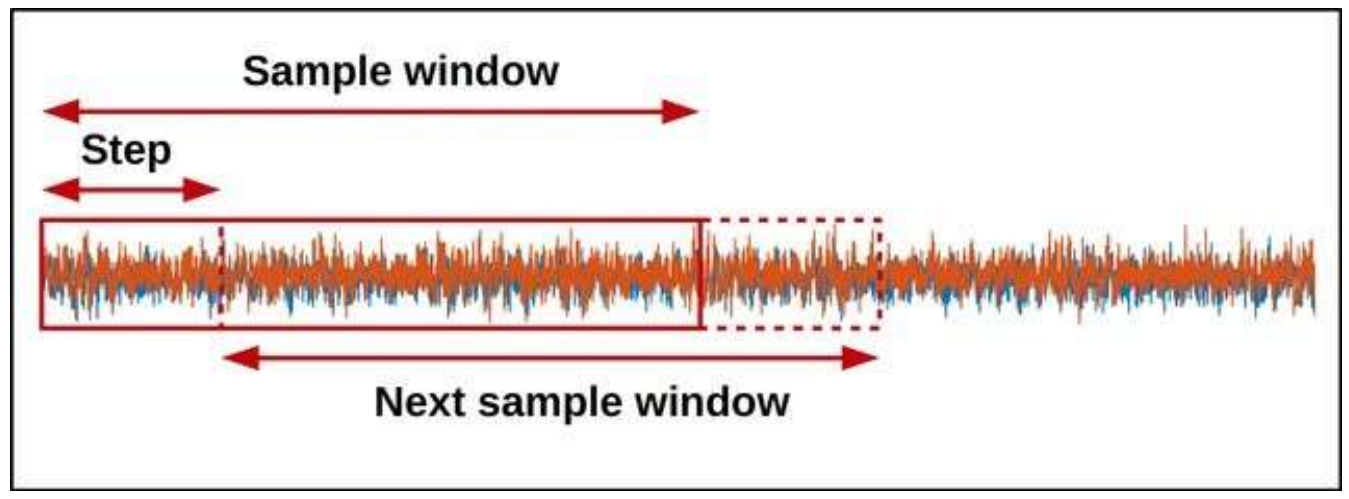

Fig. 7. Step-wise augmentation using a sliding sample window over one-second segments of the signal.

### C. Implementation and Training Details

Seventy percent of the data is used for training, fifteen percent for validation and fifteen percent for testing. Throughout the experiments the number of training steps is fixed at 100 and the learning rate at 0.01 so that the model reflects similarities appropriately. The learning rate governs the velocity of gradient descent: a value that is too high accelerates the learning trajectory at the expense of peak accuracy, whereas a value that is too low traps optimization in a local minimum and forces
longer training cycles, so the chosen value of 0.01 balances convergence speed and stability. As training progresses the rate of change gradually slows, and excessive training risks over-adaptation and a corresponding drop in training accuracy.

Overfitting is among the most important phenomena to avoid when training a transfer-learning model, as it results in a model that fits the training data well but fails to generalize to new data. Cross-validation is therefore employed: the data are split into a training set and a validation set, and the model's performance is measured on the validation set only after training. The metrics `loss` and `accuracy` quantify the training set, while `val_loss` and `val_accuracy` quantify the validation set. Fine-tuning further reduces development time by initializing the network with weights from an existing detector, so that the generic convolutional features are reused while only the final classification layer is specialized to the real/fake task. The principal implementation and training settings are listed in Table I.

### D. Evaluation Metrics

Four standard metrics are used to assess the proposed framework: accuracy, precision, recall and the F1-score.

Accuracy [32], [33] is the proportion of correctly classified samples—both fake and real—among all samples, and provides an overall measure of how well the model distinguishes manipulated content from authentic content:

$$\text{Accuracy} = \frac{TP + TN}{TP + TN + FP + FN}. \tag{1}$$

Because accuracy alone can be misleading on an imbalanced dataset such as FakeAVCeleb, where one class dominates, it is complemented by precision, recall and the F1-score.

TABLE I
IMPLEMENTATION AND TRAINING DETAILS

| Detail | Setting |
|---|---|
| Audio feature extractors | LipNet, DeepSpeech2 |
| Visual feature extractors | BlazeFace, ResNet18 |
| Face crop / resize | $299 \times 299$ pixels |
| Audio segment length | 1-second spectrograms |
| Augmentation | Sliding window over 1-s segments |
| Core classifier | Random Forest (ensemble) |
| MLP structure | Input + 2 hidden + output layers |
| LSTM sequence length | 1 |
| Loss function (MLP) | Cross-entropy |
| Optimisation | Gradient descent |
| Train / Validation / Test | 70% / 15% / 15% |
| Training steps | 100 |
| Learning rate | 0.01 |
| Datasets | FakeAVCeleb, DFDC |

Precision [32], [34] is the ratio of correctly predicted positive observations to all predicted positives and is appropriate when the cost of false positives is high:

$$\text{Precision} = \frac{TP}{TP + FP}. \tag{2}$$

Recall [32], [35] is the ratio of correctly predicted positives to all actual positives and matters when the cost of false negatives is high:

$$\text{Recall} = \frac{TP}{TP + FN}. \tag{3}$$

The F1-score [35], [36] is the harmonic mean of precision and recall and is preferable when the class distribution is non-uniform:

$$\text{F1-score} = \frac{2 \times (\text{Recall} \times \text{Precision})}{\text{Recall} + \text{Precision}}. \tag{4}$$

In all of the above formulations, the positive class is taken to be fake content and the negative class to be real content. True positives (TP) and true negatives (TN) are, respectively, the fake and real instances that the model predicts correctly, whereas false positives (FP) and false negatives (FN) are the corresponding incorrect predictions—a real sample wrongly flagged as fake, or a fake sample wrongly accepted as real. These four quantities form the basis of all the metrics defined above.

## V. EXPERIMENTAL RESULTS AND DISCUSSION

### A. Performance on the DFDC Dataset

The results from the DFDC dataset are not reported in detail because the model achieved only 52 percent accuracy on it. This dataset has several limitations, primarily inadequate labelling for the present task: while it labels the authenticity of the video, it lacks labels for audio detection and does not distinguish whether the person or the audio component is authentic.

TABLE II
RESULTS OF AUDIO FEATURES WITHOUT AUGMENTATION

| Method | Acc. | Prec. | Recall | F1 |
|---|---|---|---|---|
| **Random Forest** | 0.8615 | 0.9668 | 0.7927 | 0.8712 |
| AdaBoost Classifier | 0.8034 | 0.8081 | 0.7906 | 0.7993 |
| Decision Trees | 0.7846 | 0.7509 | 0.7934 | 0.7716 |
| Na¨ıve Bayes | 0.7828 | 0.7878 | 0.7694 | 0.7785 |
| MLP Classifier | 0.7757 | 0.7509 | 0.7782 | 0.7643 |
| SVM (linear) | 0.7283 | 0.7085 | 0.7245 | 0.7164 |
| SVM (polynomial) | 0.5898 | 0.6513 | 0.5666 | 0.6060 |
| K-Nearest Neighbor | 0.5103 | 0.3063 | 0.4911 | 0.3773 |

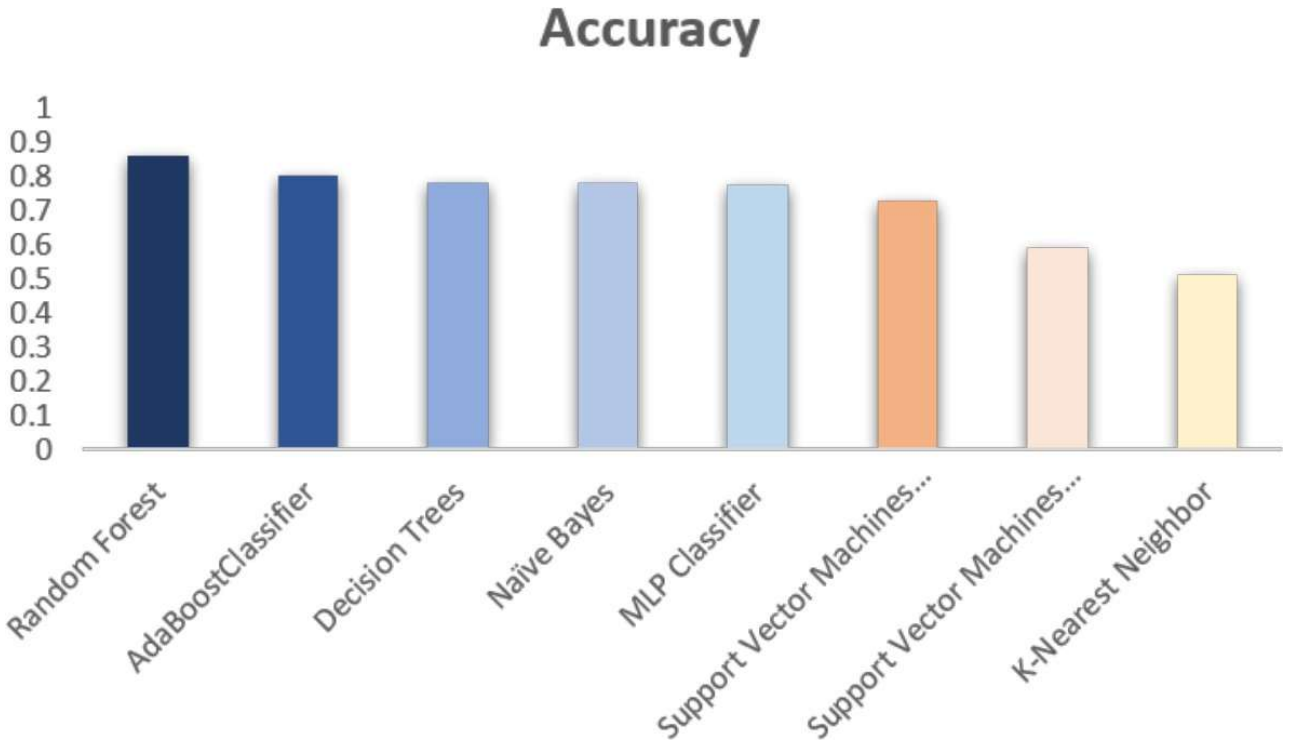

Fig. 8. Accuracy of the classification models for audio features *before* augmentation (corresponding to Table II).

TABLE III
RESULTS OF AUDIO FEATURES WITH AUGMENTATION

| Method | Acc. | Prec. | Recall | F1 |
|---|---|---|---|---|
| **Random Forest** | 0.9436 | 0.9240 | 0.9615 | 0.9424 |
| Multimodal (Ensemble) [25] | 0.8943 | 0.8318 | 0.9911 | 0.9043 |
| SVM (linear) | 0.8152 | 0.8240 | 0.8090 | 0.8164 |
| Decision Trees | 0.7424 | 0.7219 | 0.7518 | 0.7365 |
| AdaBoost Classifier | 0.7400 | 0.7288 | 0.7446 | 0.7366 |
| MLP Classifier | 0.7089 | 0.5411 | 0.8128 | 0.6497 |
| LSTM | 0.6955 | 0.6588 | 0.7299 | 0.6892 |
| Na¨ıve Bayes | 0.6392 | 0.5781 | 0.6573 | 0.6152 |
| SVM (polynomial) | 0.5794 | 0.5568 | 0.5820 | 0.5691 |
| K-Nearest Neighbor | 0.5145 | 0.3308 | 0.5210 | 0.4047 |

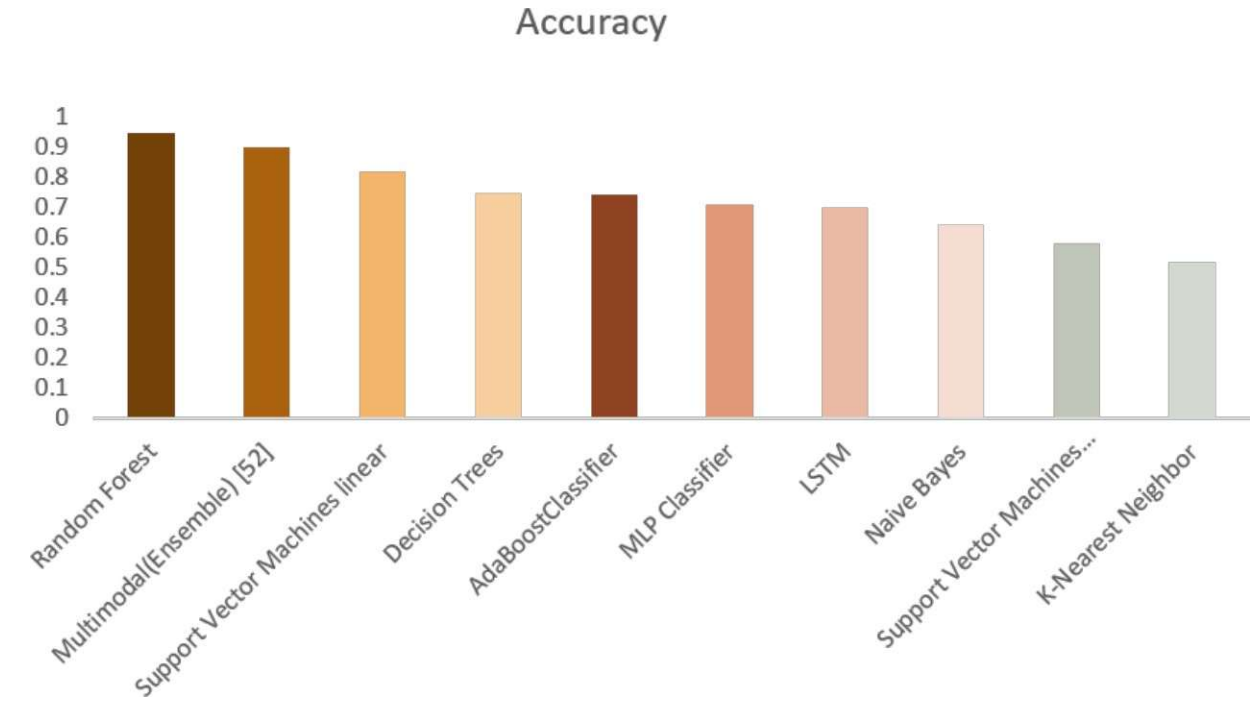

Fig. 9. Accuracy of the classification models for audio features *after* augmentation (corresponding to Table III).

### *B. Audio Detection Before Augmentation*

Without an augmented dataset, an accuracy of 86 percent is obtained for the audio part using lip-reading and speech-to-text features. The performance of all classification models on the FakeAVCeleb benchmark is compared in Table II, and the corresponding accuracy ranking is plotted in Fig. 8. Random Forest clearly leads with an accuracy of 0.8615 and the highest precision (0.9668), indicating that very few real samples are misclassified as fake. Tree-based and boosting learners (AdaBoost, Decision Trees) and the probabilistic Na¨ıve Bayes classifier form a competitive middle band around 0.78–0.80, whereas distance- and margin-based learners degrade sharply: the polynomial SVM and K-Nearest Neighbor fall to 0.59 and 0.51, respectively, the latter suffering from a precision of only 0.31. This spread confirms that the concatenated semantic feature space is high-dimensional and non-linear, conditions under which an ensemble of decision trees generalizes far better than a single distance metric.

### *C. Audio Detection After Augmentation*

With augmentation, the speech-to-text and lip-movement features yield a marked improvement. As reported in Table III and visualized in Fig. 9, Random Forest attains 94 percent accuracy, surpassing the state-of-the-art ensemble multimodal baseline [25]. Augmentation through window sliding over one-second segments enlarges the minority class and exposes the classifiers to a wider spectrum of realistic conditions, lifting the Random Forest accuracy from 0.8615 to 0.9436 and its F1-score from 0.8712 to 0.9424. Notably, the baseline ensemble achieves the highest recall (0.9911) but a lower precision (0.8318), meaning it flags almost every fake yet produces more false alarms; the proposed Random Forest instead balances precision (0.9240) and recall (0.9615) for the best overall F1-score. The remaining learners benefit from augmentation to differing degrees, but the ordering is preserved—ensemble trees dominate, while K-Nearest Neighbor remains the weakest at 0.5145, underscoring its unsuitability for this feature space.

### *D. Audio–Visual Detection After Augmentation*

When the ResNet18 visual features are combined with the audio features, no further improvement in accuracy is observed; the Random Forest accuracy settles at 0.9334, marginally below the audio-only result. This is attributed to the continual improvement of deepfake generation methods: the updated dataset contains synthetic faces that are visually indistinguishable from real ones, so neither current models nor human observers can reliably tell whether the person in the image is real or fake, as illustrated in Fig. 10. Consequently, the visual stream contributes little discriminative signal and slightly dilutes the highly informative audio features. The full audio-visual comparison is given in Table IV and Fig. 11, where Random Forest again dominates (F1-score 0.9311) and the relative ordering of the remaining classifiers is consistent with the audio-only experiments. This finding motivates the design choice of emphasizing audio and lip-sync coherence rather than relying on raw facial appearance.

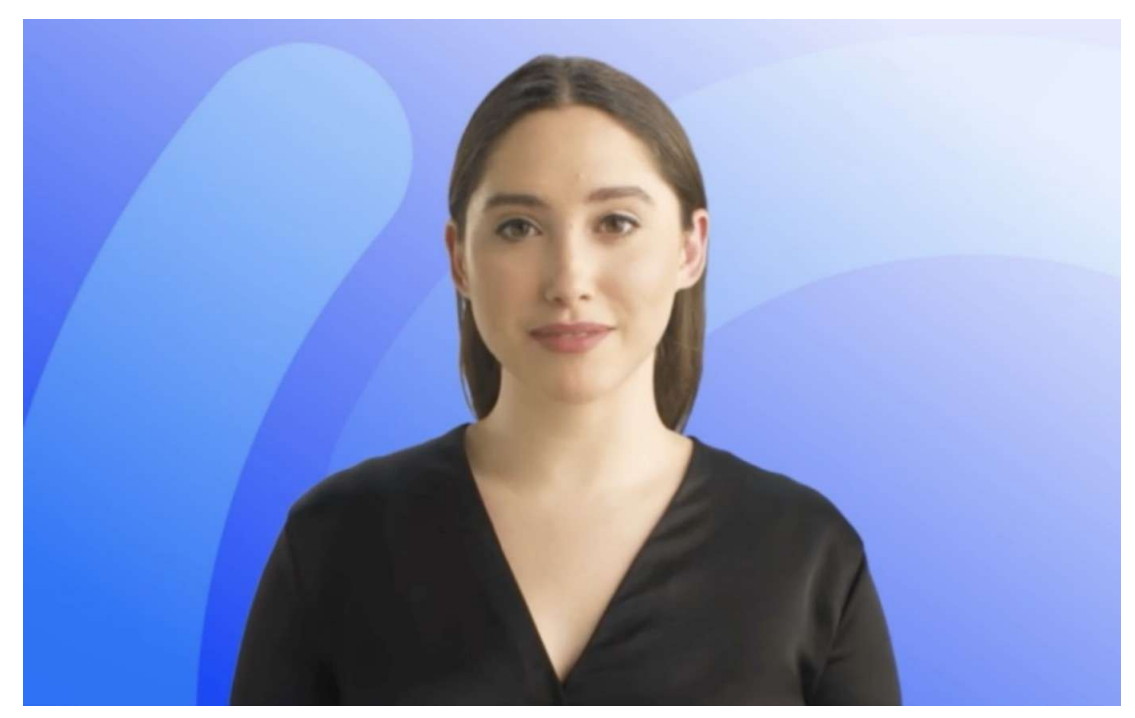

Fig. 10. Example of a synthetic (fake) person that is visually indistinguishable from a real individual.

TABLE IV
RESULTS OF AUDIO–VISUAL FEATURES AFTER AUGMENTATION

| Method | Acc. | Prec. | Recall | F1 |
|---|---|---|---|---|
| **Random Forest** | 0.9334 | 0.9077 | 0.9557 | 0.9311 |
| Decision Trees | 0.7820 | 0.8070 | 0.7659 | 0.7859 |
| AdaBoost Classifier | 0.7204 | 0.7105 | 0.7215 | 0.7160 |
| SVM (linear) | 0.6671 | 0.4218 | 0.8193 | 0.5569 |
| K-Nearest Neighbor | 0.6606 | 0.6196 | 0.6709 | 0.6442 |
| MLP Classifier | 0.6285 | 0.7264 | 0.6044 | 0.6598 |
| Naïve Bayes | 0.6244 | 0.6361 | 0.6178 | 0.6268 |
| SVM (polynomial) | 0.5605 | 0.7236 | 0.5426 | 0.6202 |

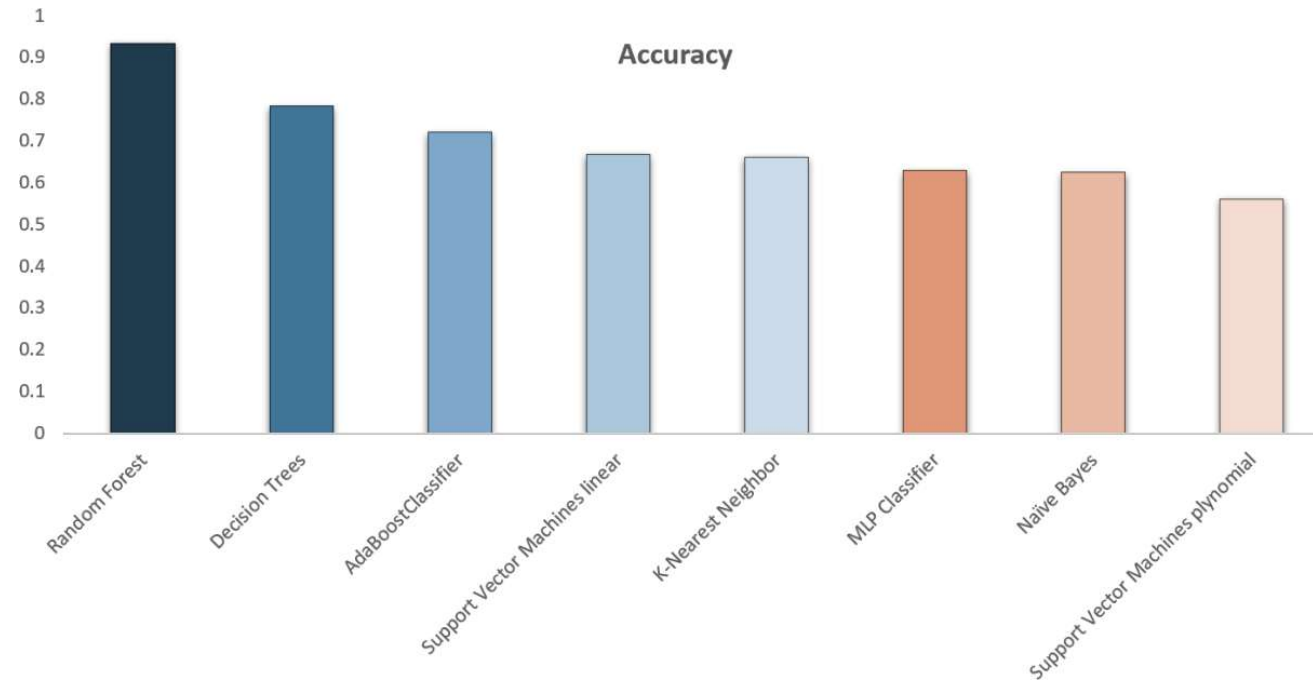

Fig. 11. Accuracy of the classification models for audio–visual features *after* augmentation (corresponding to Table IV).

### *E. Comparative Analysis*

Table V compares the proposed Random Forest based system with the state-of-the-art model across the different feature configurations. The proposed system achieves 94 percent accuracy on audio features after augmentation, exceeding the 89 percent accuracy of the state-of-the-art ensemble multi-modal model [25]. The progression 86 percent → 94 percent quantifies the benefit of augmentation, while the 93 percent obtained after adding visual features confirms that, for the current generation of high-quality deepfakes, the audio and lip-sync modalities carry the decisive evidence.

TABLE V
FINAL COMPARATIVE RESULTS (RANDOM FOREST CLASSIFIER)

| Configuration | Accuracy |
|---|---|
| State-of-the-art model [25] | 89% |
| Audio Features (Before Augmentation) | 86% |
| **Audio Features (After Augmentation)** | **94%** |
| Audio + Video Features (After Augmentation) | 93% |

### *F. Discussion*

Three consistent observations emerge from the experiments. First, the choice of classifier matters more than the addition of further modalities: across all three configurations Random Forest is the clear winner, owing to its ability to handle the high-dimensional, non-linear feature space produced by concatenating LipNet, DeepSpeech2 and ResNet18 descriptors, whereas distance-based learners such as K-Nearest Neighbor fail to separate the classes. Second, data augmentation is the single most effective lever for this imbalanced dataset, raising audio accuracy by roughly eight percentage points by enriching the minority class and simulating realistic acoustic variations. Third, the visual modality has reached a point of diminishing returns: because state-of-the-art generators now synthesize faces that are indistinguishable from real ones (Fig. 10), facial appearance no longer provides reliable discriminative cues, which explains why the audio-visual configuration does not exceed the audio-only result.

These findings also expose the limitations of the DFDC dataset for this task. Its labels indicate only whether the video is manipulated and do not specify whether the audio is authentic or whether the depicted person is real, so it could not support the fine-grained audio-visual analysis required here, which is reflected in the low 52 percent accuracy. The practical implication is that robust deepfake news detection should prioritize audio integrity and audio-visual (lip-sync) coherence, and should be trained on datasets that provide modality-specific labels. The efficiency-oriented design—retaining only the most informative sub-networks and using a lightweight Random Forest classifier—further makes the framework suitable for near-real-time screening of news content.

## VI. CONCLUSION AND FUTURE WORK

This paper presented a multimodal audio-visual framework for deepfake news detection that integrates transfer-learned features from LipNet, DeepSpeech2, BlazeFace and ResNet18 and classifies a concatenated feature vector with an ensemble of machine-learning and deep-learning models. By focusing on synchronising lip movements with audio and incorporating semantic feature analysis, the framework achieved an accuracy of 94 percent on the FakeAVCeleb dataset, outperforming state-of-the-art unimodal, ensemble and multimodal forgery detection techniques. The experiments also highlighted that, as deepfake generation continues to improve, purely visual cues become increasingly insufficient, which motivated the emphasis on audio and lip-sync coherence.

In future work, the architecture will be analysed under diverse noise perturbations and laundering scenarios to further enhance robustness in real-world conditions. The technique will be extended by training models in multiple languages to broaden its scope, and additional features that characterise an individual within a video will be explored to improve visual detection. Finally, the integration of attention mechanisms and graph neural networks will be investigated to improve both accuracy and the interpretability of the learned audio-visual representations.